\documentclass[10pt]{article}

\usepackage{url, graphicx, color, varioref, amsfonts, longtable, float,epsfig,amsthm}

\newtheorem{theorem}{Theorem}[section]
\newtheorem{lemma}[theorem]{Lemma}

\title{Quantum Searching via Entanglement and Partial Diffusion}

\author{Ahmed Younes \footnote {Birmingham, Edgbaston, B15 2TT, United Kingdom , axy@cs.bham.ac.uk .University of Alexandria, Faculty of Science, Math . \& Comp. Science Dept., ayounes2@yahoo.com  } \,\,\,\,\,\,\, 
Jon Rowe \footnote {Birmingham, Edgbaston, B15 2TT, United Kingdom , jer@cs.bham.ac.uk} 
\\ School of Computer Science \\ University of Birmingham \\
\and
Julian Miller \footnote {York, Heslington, YO10 5DD, United Kingdom, jfm@ohm.york.ac.uk}\\  Department of Electronics\\   University of York\\  
}

\begin{document}
\maketitle
\begin{abstract}
 In this paper, we will define a quantum operator that performs the inversion about the mean 
only on a subspace of the system ({\it Partial Diffusion Operator}). This operator is used in 
a quantum search algorithm that runs in $O(\sqrt{N/M})$ for searching an unstructured list of size $N$ 
with $M$ matches such that $1\le M \le N$. We will show that the performance of the algorithm 
is more reliable than known {\it fixed operators quantum search algorithms} especially for multiple matches where we can 
get a solution after a single iteration with probability over 90\% if the number of matches is approximately 
more than one-third of the search space. We will show that the algorithm will be able to handle the case 
where the number of matches $M$ is unknown in advance such that $1\le M \le N$ in $O(\sqrt{N/M})$. A performance comparison with Grover's algorithm will 
be provided.

\end{abstract}


\section{Introduction}

In 1996, Lov Grover \cite{grover96} presented an algorithm for searching an unstructured list of $N$ items 
for a single match with quadratic speed-up over classical algorithms. Grover's original algorithm exploits quantum parallelism by preparing a 
{\it uniform} superposition that represents all the items 
in the list then iterates both an oracle that marks the desired item by applying a phase shift of -1 on that item 
($e^{\underline i \theta_1}$, with $\theta_1= \pi$) and nothing on the other items ($e^{\underline i \theta_2}$, with $\theta_2= 0$)  
and an operator that performs inversion about the mean (diffusion operator) 
to amplify the amplitude of the match, the processes of this operator includes the operation $\left( {2\left| 0 \right\rangle \left\langle 0 \right| - I} \right)$ 
which applies a phase shift of -1 on the states 
within the superposition ($e^{\underline i \phi_1}$, with $\phi_1= \pi$) except the state 
$\left|0\right\rangle^{\otimes n}$ where it applies nothing 
($e^{\underline i \phi_2}$, with $\phi_2= 0$) \cite{nc00a}. To maintain consistency with literatures, 
this operation can also be 
written as $\left( {I - 2\left| 0 \right\rangle \left\langle 0 \right|} \right)$ which applies a phase shift if -1 on the state 
$\left|0\right\rangle^{\otimes n}$ ($e^{\underline i \phi_2}$, with $\phi_2= \pi$) and nothing on the other states of the 
superposition ($e^{\underline i \phi_1}$, with $\phi_1= 0$) {\it together with} a global phase shift of -1 \cite{Jozsa99}.

It was shown that the required number of iterations is approximately ${\pi}/{4} \sqrt {N}$ which is proved 
to be optimal to get the highest probability with the minimum number of iterations \cite{Zalka99}, 
such that there is only one match in the search space.

In \cite{Grover98a,Jozsa99,Gal00,long01,BK02}, Grover's algorithm is generalised by showing that the uniform 
superposition can be replaced by almost any arbitrary superposition and the phase shifts applied 
by the oracle and the diffusion operator ($e^{\underline i \theta_1},e^{\underline i\theta_2},e^{\underline i \phi_1}$ and $e^{\underline i \phi_2}$) can be generalised to deal with 
the arbitrary superposition and/or to increase the probability of success even with a factor increase in the number of iterations 
to still run in $O(\sqrt{N/M})$.
These give a larger class of algorithms for amplitude amplification using variable operators 
from which Grover's algorithm was shown to be a special case.


In another direction, work has been done trying to generalise Grover's algorithm with a uniform superposition 
for the case where there are known number of multiple matches in the search space \cite{boyer96,Chen99,Chen00a,Chen00b}, 
where it was shown that the required number of iterations is approximately 
${\pi}/{4}\sqrt {{N}/{M}}$ for small ${M}/{N}$. The required number of iterations will increase for $M>{N}/{2}$, 
i.e. the problem will be harder where it might be excepted to be easier \cite{nc00a}. Another work has been done 
for known number of multiple matches with arbitrary superposition and phase shifts \cite{Mosca98,Biron98,Brassard00,hoyer00,Li01} where the same problem 
for multiple matches occurs. In \cite{Brassard98,Mosca98,Brassard00}, a hybrid algorithm was presented to deal with this problem 
by applying Grover's fixed operators algorithm for ${\pi}/{4}\sqrt {{N}/{M}}$ times then apply one more step using different oracle 
and diffusion operator by replacing the standard phase shifts $\theta_1=\pi, \phi_1=\pi$ with accurately calculated 
phase shifts $\theta_1^{'}$ and $\phi_1^{'}$ according to the knowledge of the number of matches $M$ to get the solution with probability 
close to certainty. Using this algorithm will increase the hardware cost since we have to build one more oracle and one more  
diffusion operator for each particular $M$. For the sake of practicality, the operators should be fixed for any 
given $M$ and are able to handle the problem with high probability whether or not $M$ is known in advance.

In the case of multiple matches, where the number of matches is unknown, an algorithm for estimating the number of matches 
(known as {\it quantum counting algorithm}) was presented \cite{Brassard98,Mosca98}. In \cite{boyer96}, 
another algorithm was presented to find a match even if the number of matches is unknown which will be able to work 
if $M$ lies within the range $1\le M \le 3N/4$.
 
In this paper, we will propose a fixed operator quantum search algorithm that is able to handle the whole range $1\le M \le N$ 
more reliably whether or not the number of matches in known in advance. The plan of 
the paper is as follows: Section 2 introduces the general definition of the unstructured search problem. Section 3 
defines the partial diffusion operator \cite{Younes1}. Section 4 introduces the algorithm and an analysis on its behaviour. 
Section 5 shows a comparison with Grover's algorithm. Section 6 introduces the algorithm shown in \cite{boyer96} for 
unknown number of matches by replacing Grover's algorithm with the algorithm proposed here. The paper will end up with 
a general conclusion in Section 7.

\section{Unstructured Search Problem}

Consider an unstructured list $L$ of $N$ items. For simplicity and without loss of generality we will assume that 
$N = 2^n$ for some positive integer $n$. Suppose the items in the list are labelled with the integers $\{0,1,...,N - 1\}$,  
and consider a function (oracle) $f$ which maps an item $i \in L$ to 
either 0 or 1 according to some properties this item should satisfy, i.e. $f:L \to \{ 0,1\}$. 
The problem is to find any $i \in L$ such that $f(i) = 1$ assuming that such $i$ exists in the list. 
In conventional computers, solving this problem needs $O\left({N}/{M}\right)$ calls to the oracle (query),
where $M$ is the number of items that satisfy the oracle.

\section{Partial Diffusion}
In this section, we will define the {\it Partial Diffusion Operator} $D_p$ which performs the 
inversion about the mean only on a {\it subspace} of the system. 
The diagonal representation of the partial diffusion operator $D_p$ when applied on $n+1$ qubits system 
can take this form:

\begin{equation}
\label{ENheq13}
D_p =  \left( H^{ \otimes n}  \otimes I_1\right)\left( {2\left| 0 \right\rangle \left\langle 0 \right| - I_{n+1}} \right)\left(H^{ \otimes n}  \otimes I_1\right),
\end{equation}
\noindent
where the vector $\left| 0 \right\rangle$ used in Eqn. \ref{ENheq13} is of length $2N=2^{n+1}$, $I_k$ is the identity matrix 
of size $2^k\times 2^k$ and $H$ is the Hadamard gate $\left(H = {\textstyle{1 \over {\sqrt 2 }}}\left[ {\begin{array}{*{20}c}
   1 & {\,\,\,1}  \\
   1 & { - 1}  \\
\end{array}} \right]\right)$. 

To understand the effect of this operator, consider a general state $\left|\psi\right\rangle$ of $n+1$ qubits register:

\begin{equation}
\left| \psi  \right\rangle  = \sum\limits_{k = 0}^{2N - 1} {\delta _k \left| k \right\rangle }.
\end{equation}

For our purposes and without loss of generality, the general system $\left|\psi\right\rangle$ can be re-written as,

\begin{equation}
\label{ENheq14}
\left| \psi  \right\rangle = \sum\limits_{j = 0}^{N - 1} {\alpha _j \left( {\left| j \right\rangle  \otimes \left| 0 \right\rangle } \right)}  + \sum\limits_{j = 0}^{N - 1} {\beta _j \left( {\left| j \right\rangle  \otimes \left| 1 \right\rangle } \right)},
\end{equation}

\noindent
where \{$\alpha _j  = \delta _k$ : $k$ even\} and \{$\beta _j  = \delta _k$ : $k$ odd\}. 
The effect of applying $D_p$ on $\left| \psi  \right\rangle$ produces,

\begin{equation}
\label{ENheq15}
\begin{array}{l}
D_p\left| \psi  \right\rangle = 
\left( H^{ \otimes n}  \otimes I_1\right)\left( {2\left| 0 \right\rangle \left\langle 0 \right| - I_{n+1}} \right)\left(H^{ \otimes n}  \otimes I_1\right)
\sum\limits_{k = 0}^{2N - 1} {\delta _k \left| k \right\rangle } \\
\,\,\,\,\,\,\,\,\,\,\,\,\,\,\,\,\,\, 
= 2\left( {H^{ \otimes n}  \otimes I_1\left| 0 \right\rangle \left\langle 0 \right|H^{ \otimes n}  \otimes I_1} \right)
\sum\limits_{k = 0}^{2N - 1} {\delta _k \left| k \right\rangle }  - \sum\limits_{k = 0}^{2N - 1} {\delta _k \left| k \right\rangle } \\
\,\,\,\,\,\,\,\,\,\,\,\,\,\,\,\,\,\, 
=\sum\limits_{j = 0}^{N - 1} {2\left\langle \alpha  \right\rangle \left( {\left| j \right\rangle  \otimes \left| 0 \right\rangle } \right)}  - \sum\limits_{k = 0}^{2N - 1} {\delta _k } \left| k \right\rangle \\ 
\,\,\,\,\,\,\,\,\,\,\,\,\,\,\,\,\,\,
=\sum\limits_{j = 0}^{N - 1} {\left( {2\left\langle \alpha  \right\rangle  - \alpha _j } \right)\left( {\left| j \right\rangle  \otimes \left| 0 \right\rangle } \right)}  - \sum\limits_{j = 0}^{N - 1} {\beta _j \left( {\left| j \right\rangle  \otimes \left| 1 \right\rangle } \right)}, \\
\end{array}
\end{equation}

\noindent
where $\left\langle \alpha  \right\rangle  = \sum\nolimits_{j = 0}^{N - 1} {\alpha _j }/N$ is the mean of 
the amplitudes of the subspace entangled with the state $\left|0\right\rangle$ of the extra qubit workspace, i.e. ${\alpha _j \left( {\left| j \right\rangle  \otimes 
\left| 0 \right\rangle } \right)}$. That is, applying the operator $D_p$ will perform the inversion about the mean 
only on the subspace ${\alpha _j \left( {\left| j \right\rangle  \otimes 
\left| 0 \right\rangle } \right)}$ and will only {\it change the sign} of the amplitudes for the rest of the system 
${\beta _j \left( {\left| j \right\rangle  \otimes \left| 1 \right\rangle } \right)}$.  
A circuit implementation for $D_p$ using elementary gates \cite{elementary-gates} is shown in Fig. \ref{figY}.

\begin{center}
\begin{figure}  [t]
\begin{center}
\setlength{\unitlength}{3947sp}%
\begingroup\makeatletter\ifx\SetFigFont\undefined%
\gdef\SetFigFont#1#2#3#4#5{%
  \reset@font\fontsize{#1}{#2pt}%
  \fontfamily{#3}\fontseries{#4}\fontshape{#5}%
  \selectfont}%
\fi\endgroup%
\begin{picture}(3525,1974)(3226,-2173)
{\color[rgb]{0,0,0}\thinlines
\put(5101,-361){\circle*{150}}
}%
{\color[rgb]{0,0,0}\put(5101,-811){\circle*{150}}
}%
{\color[rgb]{0,0,0}\put(5101,-1561){\circle*{150}}
}%
{\color[rgb]{0,0,0}\put(3976,-511){\framebox(300,300){}}
}%
{\color[rgb]{0,0,0}\put(4276,-361){\line( 1, 0){225}}
}%
{\color[rgb]{0,0,0}\put(4501,-511){\framebox(300,300){}}
}%
{\color[rgb]{0,0,0}\put(3976,-961){\framebox(300,300){}}
}%
{\color[rgb]{0,0,0}\put(4276,-811){\line( 1, 0){225}}
}%
{\color[rgb]{0,0,0}\put(4501,-961){\framebox(300,300){}}
}%
{\color[rgb]{0,0,0}\put(3976,-1711){\framebox(300,300){}}
}%
{\color[rgb]{0,0,0}\put(4276,-1561){\line( 1, 0){225}}
}%
{\color[rgb]{0,0,0}\put(4501,-1711){\framebox(300,300){}}
}%
{\color[rgb]{0,0,0}\put(4801,-361){\line( 1, 0){600}}
}%
{\color[rgb]{0,0,0}\put(4801,-811){\line( 1, 0){600}}
}%
{\color[rgb]{0,0,0}\put(4801,-1561){\line( 1, 0){600}}
}%
{\color[rgb]{0,0,0}\put(4951,-2161){\framebox(300,300){}}
}%
{\color[rgb]{0,0,0}\put(5101,-361){\line( 0,-1){600}}
}%
{\color[rgb]{0,0,0}\put(5101,-1411){\line( 0,-1){450}}
}%
{\color[rgb]{0,0,0}\put(5401,-511){\framebox(300,300){}}
}%
{\color[rgb]{0,0,0}\put(5401,-961){\framebox(300,300){}}
}%
{\color[rgb]{0,0,0}\put(5401,-1711){\framebox(300,300){}}
}%
{\color[rgb]{0,0,0}\put(5401,-2161){\framebox(300,300){}}
}%
{\color[rgb]{0,0,0}\put(5926,-511){\framebox(300,300){}}
}%
{\color[rgb]{0,0,0}\put(5926,-961){\framebox(300,300){}}
}%
{\color[rgb]{0,0,0}\put(5926,-1711){\framebox(300,300){}}
}%
{\color[rgb]{0,0,0}\put(5701,-1561){\line( 1, 0){225}}
}%
{\color[rgb]{0,0,0}\put(6226,-1561){\line( 1, 0){ 75}}
}%
{\color[rgb]{0,0,0}\put(5701,-811){\line( 1, 0){225}}
}%
{\color[rgb]{0,0,0}\put(5701,-361){\line( 1, 0){225}}
}%
{\color[rgb]{0,0,0}\put(6226,-811){\line( 1, 0){ 75}}
}%
{\color[rgb]{0,0,0}\put(6226,-361){\line( 1, 0){ 75}}
}%
{\color[rgb]{0,0,0}\put(3901,-361){\line( 1, 0){ 75}}
}%
{\color[rgb]{0,0,0}\put(3901,-811){\line( 1, 0){ 75}}
}%
{\color[rgb]{0,0,0}\put(3901,-1561){\line( 1, 0){ 75}}
}%
{\color[rgb]{0,0,0}\put(4951,-2011){\line(-1, 0){1050}}
}%
{\color[rgb]{0,0,0}\put(5251,-2011){\line( 1, 0){150}}
}%
{\color[rgb]{0,0,0}\put(5701,-2011){\line( 1, 0){600}}
}%
{\color[rgb]{0,0,0}\put(3826,-286){\line( 0,-1){1350}}
}%
{\color[rgb]{0,0,0}\put(3826,-1636){\line( 1, 0){ 75}}
}%
{\color[rgb]{0,0,0}\put(3826,-286){\line( 1, 0){ 75}}
}%

\put(4105,-1236){$\vdots$}
\put(4630,-1236){$\vdots$}
\put(5080,-1236){$\vdots$}%
\put(5530,-1236){$\vdots$}%
\put(6055,-1236){$\vdots$}%

\put(4051,-436){$H$}%
\put(4051,-886){$H$}%
\put(4051,-1636){$H$}%
\put(4576,-436){$X$}%
\put(4576,-886){$X$}%
\put(4576,-1636){$X$}%
\put(5476,-436){$X$}%
\put(5476,-886){$X$}%
\put(5476,-1636){$X$}%
\put(5026,-2086){$U$}%
\put(5476,-2086){$V$}%
\put(6001,-436){$H$}%
\put(6001,-886){$H$}%
\put(6001,-1636){$H$}%
\put(6576,-511){$U = \left[ {\begin{array}{*{20}c}
   { - 1} & 0  \\
   0 & 1  \\
\end{array}} \right]$}%
\put(6576,-1636){$V = \left[ {\begin{array}{*{20}c}
   { - 1} & 0  \\
   0 & { - 1}  \\
\end{array}} \right]$}
\put(3376,-736){$n$}%
\put(3226,-961){qubits}%
\end{picture}
\end{center}
\caption{Quantum circuit representing the partial diffusion operator $D_p$ over $n+1$ qubits.}
\label{figY}
\end{figure}
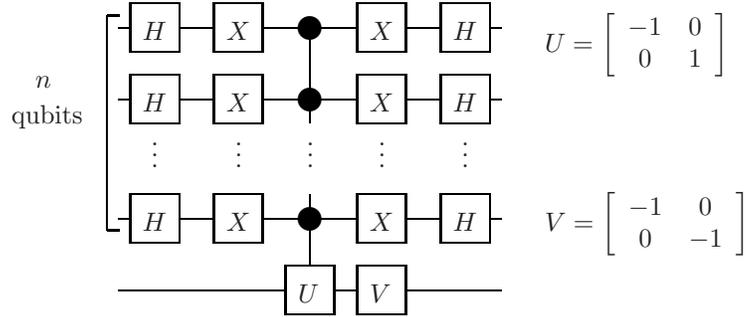
\end{center}

\section{The Algorithm}

In this section we will propose the algorithm assuming that the number of matches is known in advance. 
For a list of size $N=2^n$, prepare a quantum register of size $n+1$ 
qubits all in state $\left| 0 \right\rangle$ and apply the steps of the algorithm as follows (its quantum circuit is as shown in Fig. \ref{ENhfig6}):

\begin{itemize}

\item[1-] Apply Hadamard gate on each of the first $n$ qubits.

\item[2-]Iterate the following steps $q$ times:

\begin{itemize} 
\item[i-] Apply the oracle $U_{f}$.

\item[ii-]Apply the partial diffusion operator $D_p$. 
\end{itemize}

\item[3-] Measure the first $n$ qubits.

\end{itemize}

\begin{center}
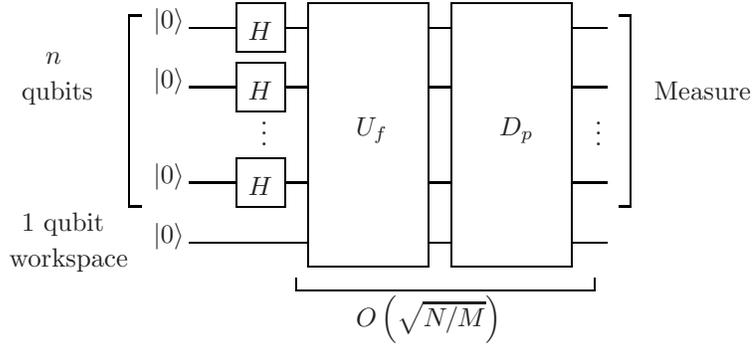
\begin{figure}  [t]
\begin{center}
\setlength{\unitlength}{3947sp}%
\begingroup\makeatletter\ifx\SetFigFont\undefined%
\gdef\SetFigFont#1#2#3#4#5{%
  \reset@font\fontsize{#1}{#2pt}%
  \fontfamily{#3}\fontseries{#4}\fontshape{#5}%
  \selectfont}%
\fi\endgroup%
\begin{picture}(4050,2091)(76,-1690)
\thinlines
{\color[rgb]{0,0,0}\put(1501, 89){\framebox(300,300){}}
}%
{\color[rgb]{0,0,0}\put(1501,-286){\framebox(300,300){}}
}%
{\color[rgb]{0,0,0}\put(1501,-886){\framebox(300,300){}}
}%
{\color[rgb]{0,0,0}\put(1951,-1261){\framebox(750,1650){}}
}%
{\color[rgb]{0,0,0}\put(2851,-1261){\framebox(750,1650){}}
}%
{\color[rgb]{0,0,0}\put(3976,-886){\line( 0, 1){1200}}
}%
{\color[rgb]{0,0,0}\put(3976,-886){\line(-1, 0){ 75}}
}%
{\color[rgb]{0,0,0}\put(3901,314){\line( 1, 0){ 75}}
}%
{\color[rgb]{0,0,0}\put(1876,-1336){\line( 0,-1){ 75}}
}%
{\color[rgb]{0,0,0}\put(1876,-1411){\line( 1, 0){1875}}
}%
{\color[rgb]{0,0,0}\put(3751,-1411){\line( 0, 1){ 75}}
}%
{\color[rgb]{0,0,0}\put(826,-886){\line( 0, 1){1200}}
}%
{\color[rgb]{0,0,0}\put(901,-886){\line(-1, 0){ 75}}
}%
{\color[rgb]{0,0,0}\put(826,314){\line( 1, 0){ 75}}
}%
{\color[rgb]{0,0,0}\put(1201,239){\line( 1, 0){300}}
}%
{\color[rgb]{0,0,0}\put(1801,239){\line( 1, 0){150}}
}%
{\color[rgb]{0,0,0}\put(2701,239){\line( 1, 0){150}}
}%
{\color[rgb]{0,0,0}\put(3601,239){\line( 1, 0){225}}
}%
{\color[rgb]{0,0,0}\put(1201,-136){\line( 1, 0){300}}
}%
{\color[rgb]{0,0,0}\put(1801,-136){\line( 1, 0){150}}
}%
{\color[rgb]{0,0,0}\put(2701,-136){\line( 1, 0){150}}
}%
{\color[rgb]{0,0,0}\put(3601,-136){\line( 1, 0){225}}
}%
{\color[rgb]{0,0,0}\put(1201,-736){\line( 1, 0){300}}
}%
{\color[rgb]{0,0,0}\put(1801,-736){\line( 1, 0){150}}
}%
{\color[rgb]{0,0,0}\put(2701,-736){\line( 1, 0){150}}
}%
{\color[rgb]{0,0,0}\put(3601,-736){\line( 1, 0){225}}
}%
{\color[rgb]{0,0,0}\put(1201,-1111){\line( 1, 0){750}}
}%
{\color[rgb]{0,0,0}\put(2701,-1111){\line( 1, 0){150}}
}%
{\color[rgb]{0,0,0}\put(3601,-1111){\line( 1, 0){225}}
}%

\put(1651,-500){$\vdots$}

\put(3751,-500){$\vdots$}

\put(301, 14){$n$}%

\put(151,-211){qubits}%

\put( 76,-1261){workspace}%

\put(151,-1036){1 qubit}%

\put(4126,-211){Measure}%

\put(980,239){$\left|0\right\rangle$}%
\put(980,-136){$\left|0\right\rangle$}%
\put(980,-736){$\left|0\right\rangle$}%
\put(980,-1111){$\left|0\right\rangle$}%

\put(1576,164){$H$}%
\put(1576,-211){$H$}%
\put(1576,-811){$H$}%

\put(3151,-436){$D_p$}%

\put(2251,-436){$U_f$}%

\put(2251,-1636){$O\left( {\sqrt {N/M} } \right)$}

\end{picture}

\end{center}
\caption{Quantum circuit for the proposed algorithm.}
\label{ENhfig6}
\end{figure}
\end{center}

\subsection{Analysis of Performance}

\begin{center}
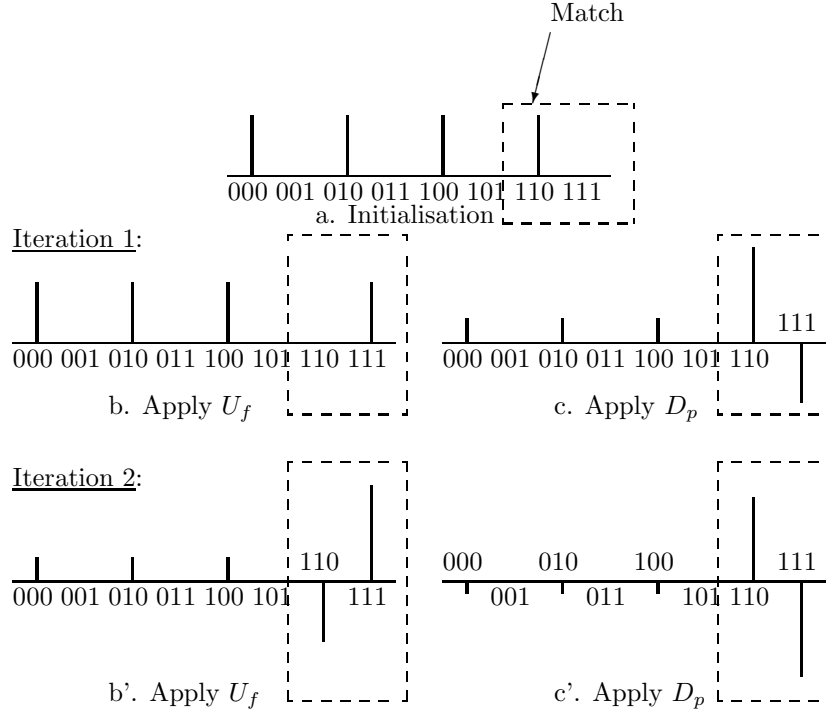
\begin{figure} [t]
\begin{center}
\setlength{\unitlength}{3947sp}%
\begingroup\makeatletter\ifx\SetFigFont\undefined%
\gdef\SetFigFont#1#2#3#4#5{%
  \reset@font\fontsize{#1}{#2pt}%
  \fontfamily{#3}\fontseries{#4}\fontshape{#5}%
  \selectfont}%
\fi\endgroup%
\begin{picture}(5199,4420)(-1511,-2501)
\thinlines
{\color[rgb]{0,0,0}\put(-149,839){\line( 1, 0){2400}}
}%
\thicklines
{\color[rgb]{0,0,0}\put(  1,1214){\line( 0,-1){375}}
}%
{\color[rgb]{0,0,0}\put(601,1214){\line( 0,-1){375}}
}%
{\color[rgb]{0,0,0}\put(1201,1214){\line( 0,-1){375}}
}%
{\color[rgb]{0,0,0}\put(1801,1214){\line( 0,-1){375}}
}%
\thinlines
{\color[rgb]{0,0,0}\put(1201,-211){\line( 1, 0){2400}}
}%
\thicklines
{\color[rgb]{0,0,0}\put(1351,-61){\line( 0,-1){150}}
}%
{\color[rgb]{0,0,0}\put(1951,-61){\line( 0,-1){150}}
}%
{\color[rgb]{0,0,0}\put(2551,-61){\line( 0,-1){150}}
}%
{\color[rgb]{0,0,0}\put(3451,-211){\line( 0,-1){375}}
}%
{\color[rgb]{0,0,0}\put(3151,389){\line( 0,-1){600}}
}%
\thinlines
{\color[rgb]{0,0,0}\put(-1499,-211){\line( 1, 0){2400}}
}%
\thicklines
{\color[rgb]{0,0,0}\put(-1349,164){\line( 0,-1){375}}
}%
{\color[rgb]{0,0,0}\put(-749,164){\line( 0,-1){375}}
}%
{\color[rgb]{0,0,0}\put(-149,164){\line( 0,-1){375}}
}%
{\color[rgb]{0,0,0}\put(751,164){\line( 0,-1){375}}
}%
\thinlines
{\color[rgb]{0,0,0}\put(-1499,-1711){\line( 1, 0){2400}}
}%
\thicklines
{\color[rgb]{0,0,0}\put(-1349,-1561){\line( 0,-1){150}}
}%
{\color[rgb]{0,0,0}\put(-749,-1561){\line( 0,-1){150}}
}%
{\color[rgb]{0,0,0}\put(-149,-1561){\line( 0,-1){150}}
}%
\thinlines
{\color[rgb]{0,0,0}\put(1201,-1711){\line( 1, 0){2400}}
}%
\thicklines
{\color[rgb]{0,0,0}\put(1351,-1711){\line( 0,-1){ 75}}
}%
{\color[rgb]{0,0,0}\put(1951,-1711){\line( 0,-1){ 75}}
}%
{\color[rgb]{0,0,0}\put(2551,-1711){\line( 0,-1){ 75}}
}%
{\color[rgb]{0,0,0}\put(3151,-1186){\line( 0,-1){525}}
}%
\thinlines
{\color[rgb]{0,0,0}\put(1876,1739){\vector(-1,-4){110.294}}
}%
{\color[rgb]{0,0,0}\put(2926,-661){\dashbox{60}(750,1125){}}
}%
\thicklines
{\color[rgb]{0,0,0}\put(751,-1111){\line( 0,-1){600}}
}%
{\color[rgb]{0,0,0}\put(451,-1711){\line( 0,-1){375}}
}%
\thinlines
{\color[rgb]{0,0,0}\put(2926,-2461){\dashbox{60}(750,1500){}}
}%
{\color[rgb]{0,0,0}\put(226,-2461){\dashbox{60}(750,1500){}}
}%
{\color[rgb]{0,0,0}\put(226,-661){\dashbox{60}(750,1125){}}
}%
\thicklines
{\color[rgb]{0,0,0}\put(3451,-1711){\line( 0,-1){600}}
}%
\thinlines
{\color[rgb]{0,0,0}\put(1576,539){\dashbox{60}(825,750){}}
}%
\put(-149,689){\makebox(0,0)[lb]{\smash{\SetFigFont{10}{12.0}{\rmdefault}{\mddefault}{\updefault}{\color[rgb]{0,0,0}000}%
}}}
\put(151,689){\makebox(0,0)[lb]{\smash{\SetFigFont{10}{12.0}{\rmdefault}{\mddefault}{\updefault}{\color[rgb]{0,0,0}001}%
}}}
\put(451,689){\makebox(0,0)[lb]{\smash{\SetFigFont{10}{12.0}{\rmdefault}{\mddefault}{\updefault}{\color[rgb]{0,0,0}010}%
}}}
\put(751,689){\makebox(0,0)[lb]{\smash{\SetFigFont{10}{12.0}{\rmdefault}{\mddefault}{\updefault}{\color[rgb]{0,0,0}011}%
}}}
\put(1051,689){\makebox(0,0)[lb]{\smash{\SetFigFont{10}{12.0}{\rmdefault}{\mddefault}{\updefault}{\color[rgb]{0,0,0}100}%
}}}
\put(1351,689){\makebox(0,0)[lb]{\smash{\SetFigFont{10}{12.0}{\rmdefault}{\mddefault}{\updefault}{\color[rgb]{0,0,0}101}%
}}}
\put(1651,689){\makebox(0,0)[lb]{\smash{\SetFigFont{10}{12.0}{\rmdefault}{\mddefault}{\updefault}{\color[rgb]{0,0,0}110}%
}}}
\put(1951,689){\makebox(0,0)[lb]{\smash{\SetFigFont{10}{12.0}{\rmdefault}{\mddefault}{\updefault}{\color[rgb]{0,0,0}111}%
}}}
\put(1201,-361){\makebox(0,0)[lb]{\smash{\SetFigFont{10}{12.0}{\rmdefault}{\mddefault}{\updefault}{\color[rgb]{0,0,0}000}%
}}}
\put(1501,-361){\makebox(0,0)[lb]{\smash{\SetFigFont{10}{12.0}{\rmdefault}{\mddefault}{\updefault}{\color[rgb]{0,0,0}001}%
}}}
\put(1801,-361){\makebox(0,0)[lb]{\smash{\SetFigFont{10}{12.0}{\rmdefault}{\mddefault}{\updefault}{\color[rgb]{0,0,0}010}%
}}}
\put(2101,-361){\makebox(0,0)[lb]{\smash{\SetFigFont{10}{12.0}{\rmdefault}{\mddefault}{\updefault}{\color[rgb]{0,0,0}011}%
}}}
\put(2401,-361){\makebox(0,0)[lb]{\smash{\SetFigFont{10}{12.0}{\rmdefault}{\mddefault}{\updefault}{\color[rgb]{0,0,0}100}%
}}}
\put(2701,-361){\makebox(0,0)[lb]{\smash{\SetFigFont{10}{12.0}{\rmdefault}{\mddefault}{\updefault}{\color[rgb]{0,0,0}101}%
}}}
\put(3001,-361){\makebox(0,0)[lb]{\smash{\SetFigFont{10}{12.0}{\rmdefault}{\mddefault}{\updefault}{\color[rgb]{0,0,0}110}%
}}}
\put(3301,-136){\makebox(0,0)[lb]{\smash{\SetFigFont{10}{12.0}{\rmdefault}{\mddefault}{\updefault}{\color[rgb]{0,0,0}111}%
}}}
\put(-899,-661){\makebox(0,0)[lb]{\smash{\SetFigFont{10}{12.0}{\rmdefault}{\mddefault}{\updefault}{\color[rgb]{0,0,0}b. Apply $U_f$}%
}}}
\put(1900,-661){\makebox(0,0)[lb]{\smash{\SetFigFont{10}{12.0}{\rmdefault}{\mddefault}{\updefault}{\color[rgb]{0,0,0}c. Apply $D_p$}%
}}}
\put(-899,-2461){\makebox(0,0)[lb]{\smash{\SetFigFont{10}{12.0}{\rmdefault}{\mddefault}{\updefault}{\color[rgb]{0,0,0}b'. Apply $U_f$}%
}}}
\put(1900,-2461){\makebox(0,0)[lb]{\smash{\SetFigFont{10}{12.0}{\rmdefault}{\mddefault}{\updefault}{\color[rgb]{0,0,0}c'. Apply $D_p$}%
}}}

\put(601,-361){\makebox(0,0)[lb]{\smash{\SetFigFont{10}{12.0}{\rmdefault}{\mddefault}{\updefault}{\color[rgb]{0,0,0}111}%
}}}
\put(301,-361){\makebox(0,0)[lb]{\smash{\SetFigFont{10}{12.0}{\rmdefault}{\mddefault}{\updefault}{\color[rgb]{0,0,0}110}%
}}}
\put(  1,-361){\makebox(0,0)[lb]{\smash{\SetFigFont{10}{12.0}{\rmdefault}{\mddefault}{\updefault}{\color[rgb]{0,0,0}101}%
}}}
\put(-299,-361){\makebox(0,0)[lb]{\smash{\SetFigFont{10}{12.0}{\rmdefault}{\mddefault}{\updefault}{\color[rgb]{0,0,0}100}%
}}}
\put(-599,-361){\makebox(0,0)[lb]{\smash{\SetFigFont{10}{12.0}{\rmdefault}{\mddefault}{\updefault}{\color[rgb]{0,0,0}011}%
}}}
\put(-899,-361){\makebox(0,0)[lb]{\smash{\SetFigFont{10}{12.0}{\rmdefault}{\mddefault}{\updefault}{\color[rgb]{0,0,0}010}%
}}}
\put(-1199,-361){\makebox(0,0)[lb]{\smash{\SetFigFont{10}{12.0}{\rmdefault}{\mddefault}{\updefault}{\color[rgb]{0,0,0}001}%
}}}
\put(-1499,-361){\makebox(0,0)[lb]{\smash{\SetFigFont{10}{12.0}{\rmdefault}{\mddefault}{\updefault}{\color[rgb]{0,0,0}000}%
}}}
\put(-1499,-1861){\makebox(0,0)[lb]{\smash{\SetFigFont{10}{12.0}{\rmdefault}{\mddefault}{\updefault}{\color[rgb]{0,0,0}000}%
}}}
\put(-1199,-1861){\makebox(0,0)[lb]{\smash{\SetFigFont{10}{12.0}{\rmdefault}{\mddefault}{\updefault}{\color[rgb]{0,0,0}001}%
}}}
\put(-899,-1861){\makebox(0,0)[lb]{\smash{\SetFigFont{10}{12.0}{\rmdefault}{\mddefault}{\updefault}{\color[rgb]{0,0,0}010}%
}}}
\put(-599,-1861){\makebox(0,0)[lb]{\smash{\SetFigFont{10}{12.0}{\rmdefault}{\mddefault}{\updefault}{\color[rgb]{0,0,0}011}%
}}}
\put(-299,-1861){\makebox(0,0)[lb]{\smash{\SetFigFont{10}{12.0}{\rmdefault}{\mddefault}{\updefault}{\color[rgb]{0,0,0}100}%
}}}
\put(  1,-1861){\makebox(0,0)[lb]{\smash{\SetFigFont{10}{12.0}{\rmdefault}{\mddefault}{\updefault}{\color[rgb]{0,0,0}101}%
}}}
\put(601,-1861){\makebox(0,0)[lb]{\smash{\SetFigFont{10}{12.0}{\rmdefault}{\mddefault}{\updefault}{\color[rgb]{0,0,0}111}%
}}}
\put(301,-1636){\makebox(0,0)[lb]{\smash{\SetFigFont{10}{12.0}{\rmdefault}{\mddefault}{\updefault}{\color[rgb]{0,0,0}110}%
}}}
\put(1501,-1861){\makebox(0,0)[lb]{\smash{\SetFigFont{10}{12.0}{\rmdefault}{\mddefault}{\updefault}{\color[rgb]{0,0,0}001}%
}}}
\put(2101,-1861){\makebox(0,0)[lb]{\smash{\SetFigFont{10}{12.0}{\rmdefault}{\mddefault}{\updefault}{\color[rgb]{0,0,0}011}%
}}}
\put(2701,-1861){\makebox(0,0)[lb]{\smash{\SetFigFont{10}{12.0}{\rmdefault}{\mddefault}{\updefault}{\color[rgb]{0,0,0}101}%
}}}
\put(3001,-1861){\makebox(0,0)[lb]{\smash{\SetFigFont{10}{12.0}{\rmdefault}{\mddefault}{\updefault}{\color[rgb]{0,0,0}110}%
}}}
\put(1201,-1636){\makebox(0,0)[lb]{\smash{\SetFigFont{10}{12.0}{\rmdefault}{\mddefault}{\updefault}{\color[rgb]{0,0,0}000}%
}}}
\put(1801,-1636){\makebox(0,0)[lb]{\smash{\SetFigFont{10}{12.0}{\rmdefault}{\mddefault}{\updefault}{\color[rgb]{0,0,0}010}%
}}}
\put(2401,-1636){\makebox(0,0)[lb]{\smash{\SetFigFont{10}{12.0}{\rmdefault}{\mddefault}{\updefault}{\color[rgb]{0,0,0}100}%
}}}
\put(3301,-1636){\makebox(0,0)[lb]{\smash{\SetFigFont{10}{12.0}{\rmdefault}{\mddefault}{\updefault}{\color[rgb]{0,0,0}111}%
}}}
\put(1876,1814){\makebox(0,0)[lb]{\smash{\SetFigFont{10}{12.0}{\rmdefault}{\mddefault}{\updefault}{\color[rgb]{0,0,0}Match}%
}}}
\put(-1499,-1111){\makebox(0,0)[lb]{\smash{\SetFigFont{10}{12.0}{\rmdefault}{\mddefault}{\updefault}{\color[rgb]{0,0,0}\underline{Iteration 2}:}%
}}}
\put(-1499,389){\makebox(0,0)[lb]{\smash{\SetFigFont{10}{12.0}{\rmdefault}{\mddefault}{\updefault}{\color[rgb]{0,0,0} \underline{Iteration 1}:}%
}}}
\put(401,539){\makebox(0,0)[lb]{\smash{\SetFigFont{10}{12.0}{\rmdefault}{\mddefault}{\updefault}{\color[rgb]{0,0,0}a. Initialisation}%
}}}
\end{picture}

\end{center}
\caption{Mechanism of amplitude amplification for the proposed algorithm with $N=4$ and $M=1$.}
\label{PdoMech}
\end{figure}
\end{center}

For the sake of clarity and to understand the behaviour of the algorithm, we will trace the algorithm during the first 
few iterations. The mechanism of amplifying the amplitudes can be understood as shown in Fig. \ref{PdoMech}. 
Now consider the algorithm if iterated once. Its behaviour can be understood as follows:

\begin{itemize}
\item[1-]{\it Register Preparation}. Prepare a quantum register of $n+1$ 
qubits all in state $\left| 0 \right\rangle $, where the extra qubit is 
used as a workspace for evaluating the oracle $U_f$, the state of the system $\left| {W_0^{(1)} } \right\rangle$ 
can be written as follows, where the subscript number refers to the step within the iteration and $(1)$ 
in the superscript refers to the iteration number:

\begin{equation}
\label{ENheq10}
\left| {W_0^{(1)} } \right\rangle = \left| 0 \right\rangle ^{ \otimes n} \otimes 
\left| 0 \right\rangle. 
\end{equation}

\item[2-] {\it Register Initialisation}. Apply Hadamard gate on each of the first $n$ qubits in parallel, 
so they contain the $2^{n}$ states representing the list, where $i$ is the integer representation of items 
in the list:

\begin{equation}
\label{ENheq11}
\left| {W_1^{(1)} } \right\rangle = {H^{ \otimes n} \otimes I} 
\,\left| {W_0^{(1)} } \right\rangle =  {\frac{1}{\sqrt N 
}\sum\limits_{i = 0}^{N - 1} {\left| i \right\rangle } }  \otimes 
\left| 0 \right\rangle.
\end{equation}

\item[3-] {\it Applying the Oracle}. Apply the oracle $U_{f}$ that maps the items in the list to either 
0 or 1 simultaneously and stores the result in the extra workspace qubit:

\begin{equation}
\label{ENheq12}
\left| {W_2^{(1)} } \right\rangle = U_f \left| {W_1^{(1)} } \right\rangle  = 
\frac{1}{\sqrt N }\sum\limits_{i = 0}^{N - 1} {\left( {\left| i 
\right\rangle \otimes \left| {f(i)} \right\rangle } \right)}. 
\end{equation}


\item[4-]{\it Partial Diffusion}. Apply the partial diffusion operator defined above. 
Let $M$ be the number of matches, which make the oracle $U_f$ evaluate to 1 (solutions) such that 
$1 \le M \le N$. Assume that $\sum\nolimits_i {{'}} $ denotes a sum over $i$ which are desired matches, 
and $\sum\nolimits_i {{''}} $ denotes a sum over $i$ which are undesired items in the list. 
So, the system  $\left| {W_2^{(1)}} \right\rangle$ shown in Eqn. \ref{ENheq12} can be written as follows:

\begin{equation}
\label{ENheq16}
\left| {W_2^{(1)}} \right\rangle  = \frac{1}{{\sqrt N }}\sum\limits_{i = 0}^{N - 1} {''\left( {\left| i \right\rangle  \otimes \left| 0 \right\rangle } \right)}  + \frac{1}{{\sqrt N }}\sum\limits_{i = 0}^{N - 1} {'\left( {\left| i \right\rangle  \otimes \left| 1 \right\rangle } \right)}. 
\end{equation}

Applying $D_p$ on $\left| {W_2^{(1)}} \right\rangle$ will result in a new system described as follows: 

\begin{equation}
\label{ENheq17}
\left| {W_3^{(1)}} \right\rangle  = a_1 \sum\limits_{i = 0}^{N - 1} {''\left( {\left| i \right\rangle  \otimes \left| 0 \right\rangle } \right)}  + b_1 \sum\limits_{i = 0}^{N - 1} {'\left( {\left| i \right\rangle  \otimes \left| 0 \right\rangle } \right)}  + c_1 \sum\limits_{i = 0}^{N - 1} {'\left( {\left| i \right\rangle  \otimes \left| 1 \right\rangle } \right)}, 
\end{equation}

\noindent
where the mean used in the definition of partial diffusion operator is,

\begin{equation}
\label{ENheq18}
\left\langle {\alpha _1 } \right\rangle  = \left( {\frac{{N - M}}{{N\sqrt N }}} \right),
\end{equation}

and $a_1$, $b_1$ and $c_1$ used in Eqn. \ref{ENheq17} are calculated as follows:

\begin{equation}
\label{ENheq19}
a_1  = 2\left\langle {\alpha _1 } \right\rangle  - \frac{1}{{\sqrt N }},\,\,\,\,\,\
b_1  = 2\left\langle {\alpha _1 } \right\rangle, \,\,\,\,\,\ 
c_1  = \frac{{ - 1}}{{\sqrt N }}.
\end{equation}

Such that,

\begin{equation}
\label{ENheq20}
\left( {N - M} \right)a_1^2  + Mb_1^2  + Mc_1^2  = 1.
\end{equation}

Notice that, the states with amplitude $b_1$ had amplitude {\it zero} before applying $D_p$ as shown in the first 
iteration in Fig. \ref{PdoMech}.

\item[5-] {\it Measurement}. If we measure the first $n$ qubits after the first iteration ($q=1$), 
then the probabilities of the system will be as follows:

\begin{itemize}
\item[i-]Probability $P_{s}^{(1)}$ to find a match out of the $M$ possible matches is given by taking into account that a 
solution $\left| i \right\rangle $ occurs {\it twice} as $\left( {\left| i \right\rangle \otimes \left| 0 
\right\rangle } \right)$ with amplitude $b_1$ and $\left( {\left| i \right\rangle \otimes \left| 1 \right\rangle 
} \right)$ with amplitude $c_1$ as shown in Eqn. \ref{ENheq17}:

\begin{equation}
\label{ENheq21}
\begin{array}{l}
 P_{s}^{(1)}  = M\left( {b_1^2  + c_1^2 } \right) \\ 
\,\,\,\,\,\,\,  = M\left( {\left( {\frac{{2\left( {N - M} \right)}}{{N\sqrt N }}} \right)^2  + \left( {\frac{{ - 1}}{{\sqrt N }}} \right)^2 } \right) \\ 
\,\,\,\,\,\,\,  = 5\left( {\frac{M}{N}} \right) - 8\left( {\frac{M}{N}} \right)^2  + 4\left( {\frac{M}{N}} \right)^3.  \\ 
 \end{array}
\end{equation}

\item[ii-] Probability $P_{ns}^{(1)}$ to find undesired result out of the states is given by:

\begin{equation}
\label{ENheq22}
P_{ns}^{(1)}  = (N - M)a_1^2. 
\end{equation}

Notice that, using Eqn. \ref{ENheq20}, 

\begin{equation}
\label{ENheq23}
P_{s}^{(1)}  + P_{ns}^{(1)}  = 1.
\end{equation}
\end{itemize}
\end{itemize}

Consider the system after first iteration shown in Eqn. \ref{ENheq17} before applying the measurement, 
the second iteration will modify the system as follows:

Applying the oracle $U_{f}$ will {\it swap} the amplitudes of 
the states which represent the matches, i.e. states with amplitudes 
$b_{1}$ will be with amplitudes $c_{1}$ and states with amplitudes $c_{1}$ will 
be with amplitudes $b_{1}$ so the system can be described as,

\begin{equation}
\label{ENheqn29}
\left| {W_1^{(2)} } \right\rangle = a_1 \sum\limits_{i = 0}^{N - 1} {''\left( 
{\left| i \right\rangle \otimes \left| 0 \right\rangle } \right)} + c_1 
\sum\limits_{i = 0}^{N - 1} {'\left( {\left| i \right\rangle \otimes \left| 
0 \right\rangle } \right)} + b_1 \sum\limits_{i = 0}^{N - 1} {'\left( 
{\left| i \right\rangle \otimes \left| 1 \right\rangle } \right)} .
\end{equation}

Applying the operator $D_p$ will change the system as follows,

\begin{equation}
\label{ENheqn30}
\left| {W_2^{(2)} } \right\rangle = a_2 \sum\limits_{i = 0}^{N - 1} {''\left( 
{\left| i \right\rangle \otimes \left| 0 \right\rangle } \right)} + b_2 
\sum\limits_{i = 0}^{N - 1} {'\left( {\left| i \right\rangle \otimes \left| 
0 \right\rangle } \right)} + c_2 \sum\limits_{i = 0}^{N - 1} {'\left( 
{\left| i \right\rangle \otimes \left| 1 \right\rangle } \right)} ,
\end{equation}

\noindent
where the mean used in the definition of partial diffusion operator is,

\begin{equation}
\label{ENheqn31}
\left\langle {\alpha _2 } \right\rangle = \frac{1}{N}\left( {\left( {N - M} 
\right)a_1 + Mc_1 } \right),
\end{equation}

\noindent
and $a_{2}$, $b_{2}$ and $c_{2}$ used in Eqn. \ref{ENheqn30} are calculated as follows:

\begin{equation}
\label{ENheqn32}
a_2 = 2\left\langle {\alpha _2 } \right\rangle - a_1 ,\,\,\,\,\,\,\,\,b_2 = 
2\left\langle {\alpha _2 } \right\rangle - c_1 ,\,\,\,\,\,\,c_2 = - b_1 ,
\end{equation}

\noindent
and the probabilities of the system are,

\begin{equation}
\label{ENheqn33a}
\begin{array}{l}
P_s^{(2)} = M\left( {b_2^2 + c_2^2 } \right) \\
\,\,\,\,\,\,\,\,\,\,\,\, = M\left( {b_2^2 + b_1^2 } \right), \\
\end{array}
\end{equation}
and,
\begin{equation}
\label{ENheqn33b}
\begin{array}{l}
P_{ns}^{(2)} = (N-M)\left( {a_2^2 } \right) \\
\,\,\,\,\,\,\,\,\,\,\,\, = (N-M)\left( {b_2 + c_2 } \right)^2 \\ 
\,\,\,\,\,\,\,\,\,\,\,\, =(N-M)\left( {b_2 - b_1 } \right)^2. \\ 
\end{array}
\end{equation}

In the same fashion, the third iteration will give the following system,

\begin{equation}
\label{ENheqn34}
\begin{array}{l}
\left| {W_1^{(3)} } \right\rangle = U_f \left| {W_2^{(2)} } \right\rangle \\
\,\,\,\,\,\,\,\,\,\,\,\,\,\,\,\,\,\,\,\,=  a_2 
\sum\limits_{i = 0}^{N - 1} {''\left( {\left| i \right\rangle \otimes \left| 
0 \right\rangle } \right)} + c_2 \sum\limits_{i = 0}^{N - 1} {'\left( 
{\left| i \right\rangle \otimes \left| 0 \right\rangle } \right)} + b_2 
\sum\limits_{i = 0}^{N - 1} {'\left( {\left| i \right\rangle \otimes \left| 
1 \right\rangle } \right)}.\\ 
\end{array}
\end{equation}

\begin{equation}
\label{ENheqn35}
\begin{array}{l}
\left| {W_2^{(3)} } \right\rangle = D_p\left| {W_1^{(3)} } \right\rangle \\
\,\,\,\,\,\,\,\,\,\,\,\,\,\,\,\,\,\,\,\, = a_3 
\sum\limits_{i = 0}^{N - 1} {''\left( {\left| i \right\rangle \otimes \left| 
0 \right\rangle } \right)} + b_3 \sum\limits_{i = 0}^{N - 1} {'\left( 
{\left| i \right\rangle \otimes \left| 0 \right\rangle } \right)} + c_3 
\sum\limits_{i = 0}^{N - 1} {'\left( {\left| i \right\rangle \otimes \left| 
1 \right\rangle } \right)} ,
\end{array}
\end{equation}

\noindent
where the mean used in $D_p$ is,

\begin{equation}
\label{ENheqn36}
\left\langle {\alpha _3 } \right\rangle = \frac{1}{N}\left( {\left( {N - M} 
\right)a_2 + Mc_2 } \right),
\end{equation}

\noindent
and $a_{3}$, $b_{3}$ and $c_{3}$ used in Eqn. \ref{ENheqn35} are calculated as follows:

\begin{equation}
\label{ENheqn37}
a_3 = 2\left\langle {\alpha _3 } \right\rangle - a_2 ,\,\,\,\,\,\,\,\,b_3 = 
2\left\langle {\alpha _3 } \right\rangle - c_2 ,\,\,\,\,\,\,c_3 = - b_2,
\end{equation}

\noindent
and the probabilities of the system are,

\begin{equation}
\label{ENheqn38a}
\begin{array}{l}
P_s^{(3)} = M\left( {b_3^2 + c_3^2 } \right) \\
\,\,\,\,\,\,\,\,\,\,\,\, = M\left( {b_3^2 + b_2^2 } \right), \\
\end{array}
\end{equation}
and,
\begin{equation}
\label{ENheqn38b}
\begin{array}{l}
P_{ns}^{(3)} = (N-M)\left( {a_3^2 } \right) \\
\,\,\,\,\,\,\,\,\,\,\,\, = (N-M)\left( {b_3 + c_3 } \right)^2 \\ 
\,\,\,\,\,\,\,\,\,\,\,\, =(N-M)\left( {b_3 - b_2 } \right)^2. \\ 
\end{array}
\end{equation}

\pagebreak
In general, the system after $q \ge 2$ iterations can be described using the 
following recurrence relations, 

\begin{equation}
\label{ENheqn39}
\left| {W^{(q)}} \right\rangle = a_q \sum\limits_{i = 0}^{N - 1} {''\left( 
{\left| i \right\rangle \otimes \left| 0 \right\rangle } \right)} + b_q 
\sum\limits_{i = 0}^{N - 1} {'\left( {\left| i \right\rangle \otimes \left| 
0 \right\rangle } \right)} + c_q \sum\limits_{i = 0}^{N - 1} {'\left( 
{\left| i \right\rangle \otimes \left| 1 \right\rangle } \right)}, 
\end{equation}

\noindent
where the mean to be used in the definition of the partial diffusion 
operator is as follows: Let $y = 1 - {M}/{N}$ and $s = {1}/{\sqrt N }$, then,

\begin{equation}
\label{ENheqn40}
\left\langle {\alpha _q } \right\rangle =  {ya_{q - 1} + (1 - y)c_{q - 
1} },
\end{equation}

\noindent
and $a_{q}$, $b_{q}$ and $c_{q}$ used in Eqn. \ref{ENheqn39} are calculated as follows:

\begin{equation}
\label{ENheqn41}
a_0 = s,\,\,a_1 = s\left( {2y - 1} \right),\,\, a_q = 2\left\langle {\alpha _q } \right\rangle - a_{q - 1},
\end{equation}

\begin{equation}
\label{ENheqn42}
b_0 =s,\,\,b_1 = 2sy,\,\,b_q = 2\left\langle {\alpha _q } \right\rangle - c_{q - 1},
\end{equation}

\begin{equation}
\label{ENheqn43}
c_0 = 0,\,\,c_1 = - s,\,\,c_q = - b_{q - 1},
\end{equation}

\noindent
and the probabilities of the system are,

\begin{equation}
\label{ENheqn44}
\begin{array}{l}
P_s^{(q)} = M\left( {b_q^2 + c_q^2 } \right) \\
\,\,\,\,\,\,\,\,\,\,\,\,= M\left( {b_q^2 + b_{q - 1}^2 } \right),\\
\end{array}
\end{equation}
and
\begin{equation}
\label{ENheqn45}
\begin{array}{l}
P_{ns}^{(q)} = (N-M)\left( {a_q^2 } \right) \\
\,\,\,\,\,\,\,\,\,\,\,\,= (N-M)\left( {b_q + c_q } \right)^2 \\
\,\,\,\,\,\,\,\,\,\,\,\,= (N-M)\left( {b_q - b_{q - 1} } \right)^2.\\
\end{array}
\end{equation}

Solving the above recurrence relations for $a_q$, $b_q$ and $c_q$ shown 
Eqn. \ref{ENheqn41}, Eqn. \ref{ENheqn42} and Eqn. \ref{ENheqn43} respectively, the closed forms are as follows:

\begin{equation}
\label{ENheqn53}
a_q = s\left( {U_q(y) - U_{q - 1}(y) } \right),\,\, b_q = sU_q(y),\,\, c_q = - sU_{q - 1}(y),
\end{equation}

\noindent
where $y = \cos \left( \theta \right) = 1-M/N$, $0 < \theta \le {\pi }/{2}$ and $U_q(y)$ is the Chebyshev polynomial of the 
second kind \cite{ChebPoly} \footnote{To clear ambiguity, $U_f$ represents the oracle function and $U_q(y)$ is the Chebyshev 
polynomial of the second kind.}which is defined as follows, 

\begin{equation}
\label{ENheqn52}
U_q \left( y \right) = \frac{\sin \left( {\left( {q + 1} \right)\theta } 
\right)}{\sin \left( \theta \right)}.
\end{equation}

The probabilities of the system,

\begin{equation}
\label{ENheqn56}
P_s^{(q)} = (1 - \cos \left( \theta \right))\left( {U_q^2(y) + U_{q - 1}^2(y) } 
\right),
\end{equation}
and,
\begin{equation}
\label{ENheqn57}
P_{ns}^{(q)} = \cos \left( \theta \right)\left( {U_q(y) - U_{q - 1}(y) } 
\right)^2.
\end{equation}

\pagebreak
Such that,

\begin{equation}
\label{ENheqn58}
\begin{array}{l}
 P_s^{(q)} + P_{ns}^{(q)} = M\left( {b_q^2 + c_q^2 } \right) + \left( {N - 
M} \right)a_q^2 \\ 
\,\,\,\,\,\,\,\,\,\,\,\,\,\, = N\left( {b_q^2 + c_q^2 } 
\right) + 2\left( {N - M} \right)c_q b_q \\ 
\,\,\,\,\,\,\,\,\,\,\,\,\,\, = \frac{1}{\sin ^2\left( \theta 
\right)}\left( {\sin ^2\left( {\left( {q + 1} \right)\theta } \right) + \sin 
^2\left( {q\theta } \right) - 2\cos \left( \theta \right)\sin \left( {\left( 
{q + 1} \right)\theta } \right)\sin \left( {q\theta } \right)} \right) \\ 
 \,\,\,\,\,\,\,\,\,\,\,\,\,\, = \frac{1}{\sin ^2\left( \theta 
\right)}\left( {\cos ^2\left( {q\theta } \right)\sin ^2\left( \theta \right) 
- \sin ^2\left( {q\theta } \right)\cos ^2\left( \theta \right) + \sin 
^2\left( {q\theta } \right)} \right) \\ 
 \,\,\,\,\,\,\,\,\,\,\,\,\,\, = \frac{1}{\sin ^2\left( \theta 
\right)}\left( {\left( {1 - \sin ^2\left( {q\theta } \right)} \right)\sin 
^2\left( \theta \right) - \sin ^2\left( {q\theta } \right)\left( {1 - \sin 
^2\left( \theta \right)} \right) + \sin ^2\left( {q\theta } \right)} \right) 
\\ 
 \,\,\,\,\,\,\,\,\,\,\,\,\,\, = \frac{\sin ^2\left( \theta 
\right)}{\sin ^2\left( \theta \right)} = 1. \\ 
 \end{array}
\end{equation}

Now, we have to calculate how many iterations, $q$, are required to find any 
match with probability close to certainty for different cases of $1 \le M 
\le N$. To find a match with high probability on any measurement, then $P_s^{(q)} $ must be 
as close as possible to one. To calculate the number of iterations, 
$q$, required to satisfy this condition, we need the following theorem.

\begin{theorem} 

Consider the following relation,

\begin{equation}
\label{ENheqn59}
P_s^{(\overline q )}=(1 - \cos \left( \theta \right))\left( { \left({U_{\overline q}\left( y \right)}\right)  ^2 + \left({U_{\overline q -1 }\left( y \right)}\right)  ^2 } \right) = 1,
\end{equation}

\noindent
where $U_{\overline q} \left( y \right)$ is the Chebyshev polynomial of the second kind, $y 
= \cos \left( \theta \right)$ and $0 < \theta \le {\pi }/{2}$, then, 

\[
\overline q  = \frac{\pi - \theta }{2\theta } \mbox{ or } \theta = \frac{\pi }{2}.
\]

\begin{proof}

From the definition of $U_{\overline q} \left( y \right)$ shown in Eqn. \ref{ENheqn52} then Eqn. \ref{ENheqn59} can take this 
form,

\[
\left( {1 - \cos \left( \theta \right)} \right)\left( {\frac{\sin ^2\left( 
{\left( {\overline q  + 1} \right)\theta } \right)}{\sin ^2\left( \theta \right)} + 
\frac{\sin ^2\left( {\overline q \theta } \right)}{\sin ^2\left( \theta \right)}} 
\right) = 1,
\]

\noindent
or,

\[
\sin ^2\left( {\left( {\overline q  + 1} \right)\theta } \right) + \sin ^2\left( 
{\overline q \theta } \right) = 1 + \cos \left( \theta \right).
\]

Using simple trigonometric identities, the above relation may take the form,

\[
\cos \left( {2\overline q \theta + 2\theta } \right) + \cos \left( {2\overline q \theta } \right) 
+ 2\cos \left( \theta \right) = 0.
\]

Using the addition formulas for cosine we get,

\[
2\cos \left( {2\overline q \theta } \right)\cos ^2\left( \theta \right) - 2\cos \left( 
\theta \right)\sin \left( {2\overline q \theta } \right)\sin \left( \theta \right) + 
2\cos \left( \theta \right) = 0,
\]

\[
2\cos \left( \theta \right)\left( {\cos \left( {2\overline q \theta } \right)\cos 
\left( \theta \right) - \sin \left( {2\overline q \theta } \right)\sin \left( \theta 
\right) + 1} \right) = 0,
\]

\[
\cos \left( \theta \right)\left( {\cos \left( {2\overline q \theta + \theta } \right) + 
1} \right) = 0.
\]

From the last equation we get,
\[
\cos \left( \theta \right) = 0 \mbox{ or } \cos \left( {2\overline q \theta + \theta } \right) = -1,
\]

\noindent
which gives the required conditions, $ \theta = {\pi }/{2} \mbox{ or } \overline q  = {(\pi - \theta )}/{2\theta }$. 

\end{proof}
\end{theorem}


The number of iterations must be 
integer, let $q = \left\lfloor \pi /2\theta\right\rfloor$ where $\left| {q  - \overline q} \right| \le 1/2$. 
And since, $\cos\left( \theta  \right) = 1- {M}/{N}$, we have $\theta  \ge \sin \left( \theta \right) 
= \sqrt{2NM - M^2 } /N$, then,

\begin{equation}
\label{ENheqn60a}
q = \left\lfloor {\frac{\pi }{{2\theta }}} \right\rfloor \le  {\frac{\pi }{{2\theta }}} \le \frac{\pi }{2\sqrt{2}}\sqrt {\frac{N}{M}}  = O\left( {\sqrt {\frac{N}{M}} } \right),
\end{equation}
where $\left\lfloor {\,\,} \right\rfloor$ is the floor operation. To determine the lower bound of $P_s^{(q)}$ 
using $q$, let $P_s^{(q)}$ to take the form,

\begin{equation}
\label{ENheqn60}
\begin{array}{l}
 P_s^{(q)}  = \frac{{1 - \cos \left( \theta  \right)}}{{\sin ^2 \left( \theta  \right)}}\left( {\sin ^2 \left( {\left( {q + 1} \right)\theta } \right) + \sin ^2 \left( {q\theta } \right)} \right) \\ 
 \,\,\,\,\,\,\,\,\,\,\,\,\, = \frac{1}{{1 + \cos \left( \theta  \right)}}\left( {1 - \cos \left( \theta  \right)\cos \left( {2q\theta  + \theta } \right)} \right). \\ 
 \end{array}
\end{equation}
We have, 
\[
\left| {q - \overline q } \right| \le \frac{1}{2},
\]
then, 
\[
\left| {\left( {2q + 1} \right)\theta  - \left( {2\overline q  + 1} \right)\theta } \right| \le \theta,
\]
and from the definition of $\overline q$,  
\[
\left( {2\overline q  + 1} \right)\theta  = \pi,
\]
then, 
\[
\cos \left( {\left( {2q + 1} \right)\theta  - \pi } \right) \le \cos \left( \theta  \right),
\]
or,
\[- \cos \left( {\left( {2q + 1} \right)\theta } \right) \le \cos \left( \theta  \right).\] 
Using this in Eqn.\ref{ENheqn60}, we get the following lower bound,

\begin{equation}
\label{ENheqn60b}
P_s^{(q)}  \ge \frac{{1 + \cos ^2 \left( \theta  \right)}}{{1 + \cos \left( \theta  \right)}} = \frac{{1 + \left( {1 - \frac{M}{N}} \right)^2 }}{{1 + \left( {1 - \frac{M}{N}} \right)}}.
\end{equation}

The minimum of the lower bound is $2\sqrt 2  - 2$ ($\approx 0.83$) when $M/N=2-\sqrt{2}$ ($\approx 0.5857$). Notice that, 
when $M/N\approx 0.5857$, the probability of success is 98.78\% after a single iteration using Eqn. \ref{ENheq21}. This minimum of the lower bound can 
be neglected with respect to the real behaviour of the algorithm.

\begin{figure}[t]
\centerline{\includegraphics[width=4.00in,height=3.0in]{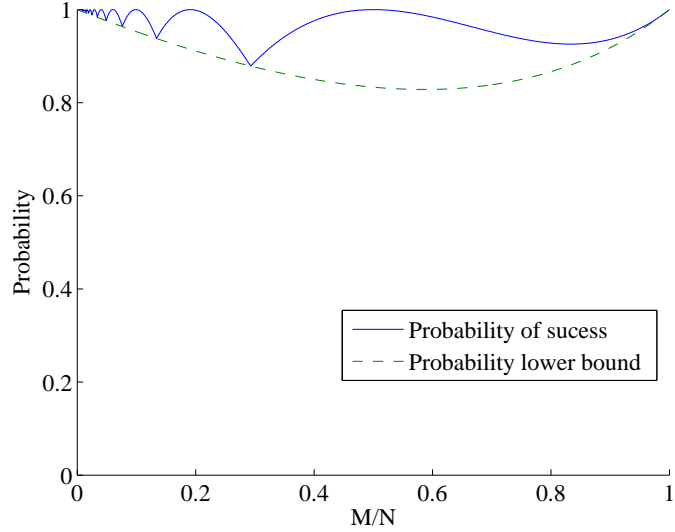}}
\caption{Probability of success of the algorithm using the required number of iterations $q$.} 
\label{pdops}
\end{figure}

To demonstrate the real behaviour of the algorithm, we may plot the 
probability of success $P_s^{(q)}$ using the required number of iterations $q$ for any given $M$, 
Fig. \ref{pdops} shows this behaviour as a function of $0<{M}/{N}\le 1$. We can see from the plot that the minimum probability 
that the algorithm may reach is approximately 87.88{\%} when $M/N \approx 0.2928$. For $0.2928<M/N\le1$, $q=1$ where 
only a single iteration is required to handle this range. For $M/N<0.2928$, $q>1$ where the algorithm will behave 
more reliable than Grover's algorithm as we will see.

\section{Comparison with Grover's Algorithm}

First we will summarise the probabilities of success and the required number of iterations for Grover's algorithm and 
the proposed algorithm before giving the comparison. The probability of success of Grover's algorithm 
as shown in \cite{boyer96} is as follows:

\begin{equation}
\label{ENheqn61}
P^{(q_G)}_{s} = \sin ^2\left( {\left( {2q_G + 1} \right)\theta_G} \right) \ge 1-\frac{M}{N},
\end{equation}

\noindent
where $\sin ^2\left( \theta_G \right) = {M}/{N},\,\,\,0 < \theta_G \le {\pi }/{2}$ and the required number of iterations $q_G$ is,

\begin{equation}
\label{ENheqn62}
q_G = \left\lfloor {\frac{\pi }{4\theta_G}} \right\rfloor \le \frac{\pi }{4}\sqrt {\frac{N}{M}}.
\end{equation}

For the proposed algorithm, the probability of success is as follows,

\begin{equation}
\label{ENheqn63}
P_s^{(q)} = \left( {1 - \cos \left( \theta \right)} \right)\left( 
{\frac{\sin ^2\left( {\left( {q + 1} \right)\theta } \right)}{\sin ^2\left( 
\theta \right)} + \frac{\sin ^2\left( {q\theta } \right)}{\sin ^2\left( 
\theta \right)}} \right),
\end{equation}

\noindent
where $\cos \left( \theta \right) = 1-{M}/{N}, \,\,\,0 < \theta \le {\pi }/{2}$ and 
the required $q$ is,
\begin{equation}
\label{ENheqn64}
q = \left\lfloor {\frac{\pi }{2\theta}} \right\rfloor \le \frac{\pi }{2\sqrt{2}}\sqrt {\frac{N}{M}} .
\end{equation}

\begin{figure}[t]
\centerline{\includegraphics[width=5.0in,height=2.2in]{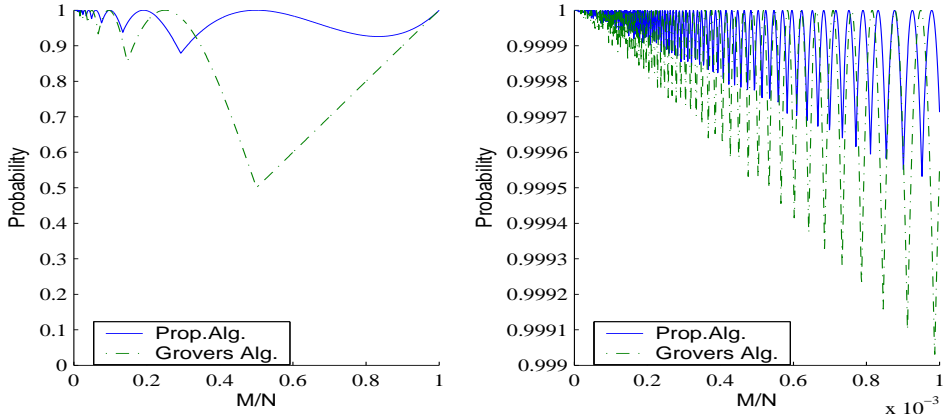}}
\caption{Probability of success using the $q_G$ and $q$ for both algorithms: 
a. $0 < {M}/{N}\le 1$ (left), b. $M / N < 1\times 10^{ - 3}$ (right).} 
\label{jitr}
\end{figure}

Fig. \ref{jitr}.a shows the probability of success for both algorithms using the required number of iterations. 
We can see from the plot that the minimum probability that Grover's algorithm may reach is approximately 
50.0{\%} when $M/N \approx 0.5$ while for the proposed algorithm, the minimum probability is 87.88{\%} 
when $M/N \approx 0.2928$. Grover's algorithm will behave similar to the single guess technique for $M/N>0.5$ where $q_G=0$ in that range so that $P^{(q_G)}_{s}=M/N$.  Although the proposed algorithm is slower than 
Grover's algorithm for small $M/N$ by $\sqrt{2}$, Fig. \ref{jitr}.b shows the probability of success for both 
algorithms for small $M/N$ (hard cases where $M / N < 1\times 10^{ -3})$, where we can see that the proposed 
algorithm is more reliable (higher probability) than Grover's algorithm. For $M/N > 1/3$, $q=1$ where the 
proposed algorithm runs in $O(1)$ to get probability at least 90\%, i.e. the problem is easier for multiple matches.

\section{Unknown Number of Matches}

In case we do not know the number of matches $M$ in advance, we can apply the algorithm shown in \cite{boyer96} 
for $1 \le M\le N$ by replacing Grover's step with the proposed algorithm. The algorithm can be summarised as follows:

\begin{itemize}
\item[1-] Start with $m=1$ and $\lambda= {8}/{7}$. (where $\lambda$ can take any value between 
1 and ${4}/{3}$)
\item[2-]Pick an integer $j$ between 0 and $m-1$ in a uniform random manner.
\item[3-]Run $j$ iterations of the proposed algorithm on the state:
 
\[
\frac{1}{{\sqrt N }}\sum\limits_{i = 0}^{N - 1} {\left| i \right\rangle \otimes \left| 0 \right\rangle}.
\]

\item[4-]Measure the register and assume $i$ is the output.
\item[5-]If $f(i)=1$, then we found a solution and exit.
\item[6-]Let $m=min\left( \lambda m,\sqrt{N}\right)$ and go to step 2.
\end{itemize}

For the sake of simplicity and to be able to compare the performance of this algorithm with that shown in \cite{boyer96}, 
we will try to follow the same style of analysis used in \cite{boyer96}. Before we construct the analysis, we need 
the following lemmas. The first lemma is straightforward using mathematical induction.

\begin{lemma}
\label{lemch5}
For any positive integer $m$ and real number $\theta$ such that $0 < \theta  \le \pi /2$, 
\[
\sum\limits_{q = 0}^{m - 1} {\sin ^2 \left( {\left( {q + 1} \right)\theta } \right) + \sin ^2 \left( {q\theta } \right) = m - \frac{{\cos \left( \theta  \right)\sin \left( {2m\theta } \right)}}{{2\sin \left( \theta  \right)}}} .
\]
\end{lemma}

\begin{lemma}

Assume $M$ is the unknown number of matches such that $1\le M \le N$. Let $\theta$ be a real number such that 
$\cos \left( \theta  \right) = 1 - M/N$ and $0 < \theta  \le \pi /2$. Let $m$ be any positive integer. 
Let $q$ be any integer picked in a uniform random manner between 0 and $m-1$. Measuring the register after applying 
$q$ iterations of the proposed algorithm starting from the initial state, the probability $P_m$ of finding a solution 
is as follows,

\[
P_m  = \frac{1}{{1 + \cos \left( \theta  \right)}}\left( {1 - \frac{{\cos \left( \theta  \right)\sin \left( {2m\theta } \right)}}{{2m\sin \left( \theta  \right)}}} \right),
\]

\noindent
where, $P_m>0.2725$ for $m \ge 1/\sin \left( \theta  \right)$ and small $M/N$.

\begin{proof} 

The average probability of success when applying $q$ iterations of the proposed algorithm when  
$0\le q \le m$ is picked in a uniform random manner is as follows,

\[ 
\begin{array}{l}
 P_m  = \sum\limits_{q = 0}^{m - 1} {\frac{1}{m}} P_s^{(q)}  = \frac{1}{{m\left( {1 + \cos \left( \theta  \right)} \right)}}\sum\limits_{q = 0}^{m - 1} {\sin ^2 \left( {\left( {q + 1} \right)\theta } \right) + \sin ^2 \left( {q\theta } \right)}  \\ 
 \,\,\,\,\,\,\,\, = \frac{1}{{1 + \cos \left( \theta  \right)}}\left( {1 - \frac{{\cos \left( \theta  \right)\sin \left( {2m\theta } \right)}}{{2m\sin \left( \theta  \right)}}} \right). \\ 
 \end{array}
 \]

\noindent
If $m \ge 1/\sin \left( \theta  \right)$ and $M \ll N$ then $ \cos\left(\theta\right)\approx 1$, so,

\[
P_m  > \frac{1}{2} - \frac{{\sin \left( {2m\theta } \right)}}{{4m\sin \left( \theta  \right)}} \ge \frac{1}{2} - \frac{{\sin \left( {2m\theta } \right)}}{4} > 0.2725,
\]

\noindent
where $\sin \left( {2m\theta } \right) < 0.91$ for $0 < \theta  \le \pi /2$. 

\end{proof}
\end{lemma}

We calculate the total expected number of iterations as done in Theorem 3 in \cite{boyer96}. Assume that 
$m_q  \ge 1/\sin \left( \theta  \right)$, and $v_q  = \left\lceil {\log _\lambda  m_q } \right\rceil $. 
Notice that, $m_q  = O\left( {\sqrt {N/M} } \right)$ for $1 \le M \le N$, then:  

\begin{itemize}

\item[1-] The total expected number of iterations to reach the critical stage, i.e. when $m\ge m_q $:

\[
\frac{1}{2}\sum\limits_{v = 1}^{v_q } {\lambda ^{v - 1} }  = \frac{1}{{2\left( {\lambda  - 1} \right)}}m_q  = 3.5m_q.
\]
 
\item[2-] The total expected number of iterations after reaching the critical stage:

\[
\frac{1}{2}\sum\limits_{u = 0}^\infty  {\left( {0.7275} \right)^u \lambda ^{v_q  + u}  = \frac{1}{{2\left( {1 - 0.7275\lambda } \right)}}} m_q  = 2.9m_q.
\]

\end{itemize}

The total expected number of iterations whether we reach to the critical stage or not is $6.4m_q$ which is in $O(\sqrt{N/M})$  for $1\le M\le N$.

\begin{figure}[t]
\centerline{\includegraphics[width=4.00in,height=3.0in]{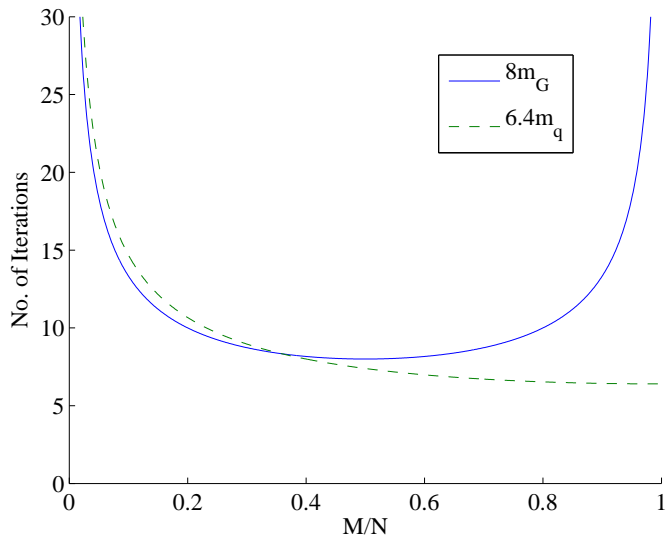}}
\caption{The actual behaviour of the functions representing the total expected number of iterations for Grover's algorithm $8 m_G$ and the 
proposed algorithm $6.4 m_q$ taking $\lambda=8/7$, where the number of iterations is the flooring of the values (step function).} 
\label{78mqg}
\end{figure}

When this algorithm employed Grover's algorithm \cite{boyer96}, and based on the condition 
$m_G  \ge 1/\sin \left( {2\theta _G } \right)= O\left( {\sqrt {N/M} } \right)$ for $1\le M\le {3N}/{4}$, where $m_G$ will act 
as a lower bound for $q_G$ in that range. The total expected number of iterations is approximately $8m_G$. 
For $M > {3N}/{4}$, $m_G$ will increase exponentially where it will not be able to approximate $q_G$. Employing the proposed algorithm instead, and based on the condition 
$m_q  \ge 1/\sin \left( {\theta} \right)= O\left( {\sqrt {N/M} } \right)$ for $1\le M\le N$,
the total expected number of iterations is approximately $6.4m_q$, i.e. the algorithm will be able to 
handle the whole range, since $m_q$ will be able to act as a lower bound for $q$ over $1\le M\le N$. 
Fig. \ref{78mqg} compares between the total expected number of iterations for both algorithms taking $\lambda  = 8/7$.

\section{Conclusion}

In Grover's algorithm, the search space is split into two subspaces 
(the solution and non-solution subspaces) then amplifies the amplitudes of the 
solution states by iterating the diffusion operator and the oracle \cite{grover96} to find a match with high probability in $O(\sqrt{N/M})$ for small $M/N$ and in the neighbourhood of $M/N=1/4$ \cite{boyer96}. 
The main idea of using partial diffusion in quantum search is to split the subspace of the solutions into two smaller subspaces. 
In each iteration, one of the solution subspaces will be inverted 
about the mean (together with the non-solution subspace) while the other half will have the sign of 
their amplitudes changed to the negative sign, preparing it to be inverted about the mean 
(together again with the non-solution subspace) in the next iteration. 
The benefit of this alternating inversion is to preserve half the number of the solution states 
at each iteration so as to resist the {\it de-amplification behaviour} of the standard diffusion 
operator when reaching the so-called turning points and get the solution with high probability in $O(\sqrt{N/M})$ for $1\le M \le N$. 
Apply the oracle $U_f$ each iteration will switch the entanglement of the two solution subspaces with the extra qubit 
workspace to decide which subspace to be inverted about the mean with the non-solution subspace.

An algorithm for unknown number of matches replacing Grover's step in the algorithm shown in  \cite{boyer96} is presented, 
where we showed that the algorithm will be able to handle the range $1\le M \le N$ in $O(\sqrt{N/M})$ compared with 
$1\le M \le 3N/4$ when using Grover's algorithm.

We showed that the algorithm will be able to handle the whole possible range $1\le M\le N$ more reliably 
using fixed operators in $O\left( {\sqrt {N/M} } \right)$ for both known and unknown number of matches which makes it 
more suitable for practical purposes.


\end{document}